\documentstyle[twoside,fleqn,espcrc2]{article}

\newcommand{\eqalign}[1]{
\null \,\vcenter {\openup \jot \ialign {\strut \hfil $\displaystyle {
##}$&$\displaystyle {{}##}$\hfil \crcr #1\crcr }}\,}
\newcommand{\be}{\begin{equation}}
\newcommand{\ee}{\end{equation}}
\newcommand{\ba}{\begin{array}}
\newcommand{\ea}{\end{array}}
\newcommand{\bea}{\begin{eqnarray}}
\newcommand{\eea}{\end{eqnarray}}
\newcommand{\pa}{\partial}
\newcommand{\scr}[1]{\scriptscriptstyle#1}

\newcommand{\AmS}{{\protect\the\textfont2
  A\kern-.1667em\lower.5
ex\hbox{M}\kern-.125emS}}

\title{Perturbative Quantum (In)equivalence
       of Dual $\sigma$ Models in $2$ dimensions}

\author{J. Balog\address{MPI,
        M\"unchen, \\
        P.O. Box 40 12 12, M\"unchen, Germany}
        P. Forg\'acs\address{D\'epartement de Physique,
        Facult\'e des Sciences et Techniques, \\
        Universit\'e de Tours, Parc de Grandmont, F-37200 Tours,
        France}
        Z. Horv\'ath and L.Palla
        \address{Institute of Theoretical Physics,
        E\"otv\"os University, \\
        H-1088, Budapest, Puskin u. 5-7, Hungary}
}

\begin{document}

\begin{abstract}
Various examples of target space duality transformations are investigated
up to two loop order in perturbation theory. Our results show that when
using the tree level (`naive') transformation rules the dual theories
are in general {\it inequivalent} at two loops to the original ones,
(both for the Abelian and the non Abelian duality).
\end{abstract}

\maketitle

\section{Introduction}
Various duality transformations
connecting two seemingly
different sigma-models
or string-backgrounds are playing an increasingly important role        
in string theories.     
It is assumed that models related by certain classical
transformations are alternative descriptions of the
same physical system (also at the quantum level).
Here we shall consider several examples of the
so called `target space duality' (Abelian T-duality) \cite{busch}, which is
the generalization
of the $R\rightarrow 1/R$ duality in toroidal compactification of string
theory. T-duality
is usually formulated in the $\sigma$-model description
of the corresponding Conformal Field Theory (CFT) (for a recent review
see \cite{review}).

It has been recently shown that
both the Abelian \cite{alglo} and the non Abelian
T-duality \cite{freeto}, \cite{fj}, \cite{frats}, \cite{curza}
transformation rules can be recovered in an elegant way by
performing a canonical transformation. This clearly shows that models
related by T-duality are {\sl classically} equivalent.
By making  some formal manipulations in the
functional integral without going, however, into the
thorny details of regularization, it is not difficult to argue
that models which are related by duality
transformations correspond to the the same Quantum Field Theory (QFT)
\cite{busch}, \cite{frats}.
For Conformal Field Theories it has been been  convincingly
argued in Ref.~\cite{rove}
that the original and the dual models are nothing but two different
functional integral representations of the same CFT.
Nevertheless it has been already pointed out in Ref.~\cite{busch2}
that the tree level transformations might
be modified by higher order terms in $\alpha'$ and the first non-trivial
correction (at the two-loop level) has been found for a special class of
$\sigma$-model in Ref.~\cite{tse1}.
Therefore we feel
that the question of quantum equivalence between 
$\sigma$-models related by duality deserves further study.

In this contribution we shall consider T-duality
transformations for $\sigma$-models, treated as `ordinary'
(i.e.~not necessarily conformally invariant)
two dimensional quantum field theories. 
More precisely we investigate the quantum equivalence of two dimensional
(2d) $\sigma$-models
related by either the Abelian \cite{busch}, or the non-Abelian
\cite{frats} version
of T-duality in the framework of perturbation theory.

Throughout the paper we work in a field
theoretic rather than string theoretic framework, that is we consider
$\sigma$-models on a flat non-dynamical 2d space.
Since the world-sheet is then flat we ignore
the dilaton completely. Furthermore only purely bosonic
$\sigma$-models shall be considered (in general also with torsion).

To
investigate the quantum equivalence of dual $\sigma$-models we compute
some `physical' quantities (up to two loops) in standard perturbation
theory in both the original and in the dual models. The perturbative
calculations are
greatly simplified if the model admits a sufficient degree
of symmetry. Therefore we have choosen models (with high enough symmetry)
where
the complete renormalization amounts to
multiplicative renormalization of the coupling(s) and then it is
not difficult to derive the corresponding $\beta$ functions.

We have investigated various Abelian duals of a `deformed'
principal $SU(2)$ $\sigma$-model and the non Abelian dual of the
principal $SU(2)$ $\sigma$-model.
In all cases we found that
up to the one loop order in perturbation theory the duals are indeed
equivalent
to the original models, though in some cases highly non trivial field
redefinitions were needed to reach this conclusion.
At the two loop
level, however, we have found that the `naive' (i.e.~tree level) duality
transformations break down and in most cases the dual models turned out to be
non renormalizable (in the restricted, field theoretic sense).
Then we could not even extract the $\beta$ functions at the two loop
level. At this point we should like to emphasize that this problem
has nothing to do with the presence or absence of the dilaton field.

A natural way to try to overcome the problem of non renormalizability
is that the duality transformation
rules for the {\sl renormalized} metric and antisymmetric tensor field
are in general modified perturbatively beyond one loop order.
This is not very surprising since in the $\sigma$-model
framework the `naive' duality relates {\sl bare} quantities \cite{bfhp}.

\section{Abelian duality}
Let us  start with a brief summary
of the Abelian T-duality \cite{busch}, \cite{rove}.
Consider the following gauged $\sigma$-model action:
\be\eqalign{
\label{gsm}
S=&\frac{1}{4\pi\alpha'}\int d^2\xi
\bigl[\sqrt{h}h^{\mu\nu}\bigl(g_{00}D_\mu D_\nu\cr
&+2g_{0\alpha}D_\mu\partial_\nu x^{\alpha} 
+g_{\alpha\beta}\partial_\mu
x^{\alpha}\partial_\nu x^{\beta}\bigr)\cr
&+i\epsilon^{\mu\nu}(2b_{0\alpha}D_\mu \partial_\nu x^{\alpha}
+b_{\alpha\beta}\partial_\mu
x^{\alpha}\partial_\nu x^{\beta})\cr
&+2i \epsilon^{\mu\nu}{\tilde \theta} \partial_\mu A_\nu\bigr]
}\ee
where $D_\mu=\partial_\mu\theta+A_\mu$,
$g_{ij}$ is the target space metric, $b_{ij}$ the
torsion, and the target space indices are decomposed as $i=(0,\alpha)$
corresponding to the coordinate decomposition $x^i=(\theta, x^\alpha)$.
The target space metric and torsion are assumed to possess a Killing
vector and are now written in the adopted coordinate system,
i.e.\ they are independent of the coordinate $\theta$.
$h_{\mu\nu}$ is the world sheet metric and
$\alpha^{'}$ the inverse of the string tension.
The $\tilde\theta$ variable is just a Lagrangian multiplier
which on topologically trivial world sheet forces
$A_\mu=\partial_\mu \epsilon$ leading to a standard (i.e.~not gauged)
$\sigma$-model, referred to as the original model.
As already alluded to in the introduction the dilaton field is ignored
in what follows and the world sheet metric, $h_{\mu\nu}$, is taken to be
flat (e.g.~that of a torus, to regulate the infrared divergences).
Then the formal functional integration over $\tilde\theta$ can be
made somewhat more precise, however, our problem is independent
of zero modes and therefore we shall not enter into more details
about them.
On the orther hand  since the action (\ref{gsm}) is quadratic
in the $A_\mu$-fields by fixing the gauge $\theta=0$ and
integrating over them one finds the dual theory:
\be\eqalign{
\label{dsm}
&{\tilde S}=\frac{1}{4\pi\alpha'}\int d^2\xi
\bigl[\sqrt{h}h^{\mu\nu}\bigl({\tilde g}_{00}\partial_\mu {\tilde \theta}
\partial_\nu {\tilde \theta}\cr
&+2{\tilde g}_{0\alpha}\partial_\mu {\tilde \theta} \partial_\nu
x^{\alpha}+{\tilde g}_{\alpha\beta}\partial_\mu
x^{\alpha}\partial_\nu
x^{\beta}\bigr)\cr
&+i\epsilon^{\mu\nu}(2{\tilde b}_{0\alpha}\partial_\mu
{\tilde \theta}\partial_\nu
x^{\alpha}+{\tilde b}_{\alpha\beta}
\partial_\mu x^{\alpha}\partial_\nu
x^{\beta})\bigr],
}\ee
where:
\be\eqalign{
\label{dmetr}
&{\tilde g}_{00}={1\over g_{00}}\cr
&{\tilde g}_{0\alpha}={b_{0\alpha} \over g_{00}},
\qquad
{\tilde b}_{0\alpha}={g_{0\alpha} \over g_{00}} \cr
&{\tilde g}_{\alpha\beta}=g_{\alpha\beta} -
{g_{0\alpha}g_{0\beta} - b_{0\alpha} b_{0\beta}\over g_{00}}\cr
&{\tilde b}_{\alpha\beta}=b_{\alpha\beta}-{g_{0\alpha}b_{0\beta}
         -g_{0\beta}b_{0\alpha}\over g_{00}}.
}\ee
These formulae (Abelian T-duality) were first found by Buscher \cite{busch}.
In this formal derivation there are, however, some hidden (potential) problems.
First there is a field-dependent determinant multiplied
by a quadratically divergent quantity (`$\delta^{(2)}(0)$') argued usually 
away by using dimensional regularization.
In the exceptionally simple case when $g_{00}$ is a constant, however, one
would expect no problems whatsoever.
As we shall show on an example (the `$\psi$-dual' model)
while the equivalence of the dual model seems indeed true,
(up to two loops in perturbation theory)
there are
nontrivial renormalization effects  even in this case
when using dimensional regularization.
It is natural to assume that the use of dimensional regularization
is part of the problem, as
the two dimensional
antisymmetric tensor, $\epsilon_{\mu\nu}$ is also present.

\section{Counterterms and the renormalization of couplings}

Our general strategy to carry out the renormalization of the
`original' and of the `dual' models
and to obtain the corresponding $\beta$ functions
is to simply use the one resp.~two
loop counterterms for the general $\sigma$-models (either with or without
the torsion term) computed by several authors \cite{hulto}, \cite{metse},
\cite{osb1}.
These counterterms  were
derived by the background field method in the dimensional
regularization scheme.
To carry out the coupling and wave function
renormalization explicitly we recall the basic formalae needed.
The general $\sigma$-model Lagrangian is written in the form
\be
\label{lag0}
{\cal L}={1\over2\lambda}\bigl( g_{ij}(\xi)+b_{ij}(\xi)\bigr)
\Xi^{ij}
={1\over\lambda}\tilde{\cal L}
\ee
where
\be
\Xi^{ij}=(\pa_\mu\xi^i\pa^\mu\xi^j
+\epsilon_{\mu\nu}\pa^\mu\xi^i\pa^\nu\xi^j)\,.
\ee
Expressing the loop expansion parameter, $\alpha^\prime$, in terms of
the coupling $\lambda$ as $\alpha^\prime=\lambda/(2\pi)$, the
simple pole parts of the one
($i=1$) and two ($i=2$) loop countertems, ${\cal L}_i$, apart from the
$\mu^{-\epsilon}$ factor are given as:
\be\label{lag11}
\mu^{\epsilon}{\cal L}_1={\alpha^\prime\over2\epsilon\lambda}
\hat{R}_{ij}\Xi^{ij}={1\over\pi\epsilon}\Sigma_1 \,.
\ee
The two loop counterterm then has the form:
\be\label{lag12}
\mu^{\epsilon}{\cal L}_2={\alpha^{\prime2}\over16\epsilon\lambda}
Y^{lmk}_{{\phantom{lmk}j}}\hat{R}_{iklm}\Xi^{ij}
={\lambda\over8\pi^2\epsilon}\Sigma_2\,,
\ee
where
\be\label{y}
\eqalign{
Y_{lmkj}=&-2\hat{R}_{lmkj}+3\hat{R}_{[klm]j}\cr
&+2(H^2_{kl}g_{mj}-H^2_{km}g_{lj})\,,\cr
H^2_{ij}=&H_{ikl}H_j^{kl}\,,\cr
2H_{ijk}=&\partial_ib_{jk}+{\rm cyclic}\,.
}\ee
In Eqs.~(\ref{lag11},\ref{lag12})
 $\hat{R}_{iklm}$ resp.\ $\hat{R}_{ij}$ denote
the `generalized' Riemann resp.\ Ricci tensors of the
`generalized' connection, $G^i_{jk}$, containing also the torsion term
in addition to the Christoffel
symbols, $\Gamma^i_{jk}$ of the metric $g_{ij}$;
\be\label{con}
G^i_{jk}=\Gamma^i_{jk}+H^i_{jk}\,.
\ee
We shall also consider examples with an additional
{\sl parameter}, $x$, where $x$ is not assumed to be small. Thus we do not
expand anything in a parameter, the standard perturbative expansion is
made only in the coupling $\lambda$.
If the metric, $g_{ij}$, and the torsion
potential, $b_{ij}$, also depend  on a (bare) parameter, $x$ i.e\
$g_{ij}=g_{ij}(\xi ,x)$ and $b_{ij}=b_{ij}(\xi ,x)$ then we convert the
previous counterterms into coupling and parameter renormalization
by assuming that in the one ($i=1$) and two ($i=2$) loop orders
their bare and renormalized values are related as
\be\label{couplren}
\eqalign{
\lambda_0&=\mu^\epsilon\lambda\Bigl( 1+{\zeta_1(x)\lambda\over\pi\epsilon}
+{\zeta_2(x)\lambda^2\over8\pi^2\epsilon}+...\Bigr)\cr
&=\mu^\epsilon\lambda Z_{\lambda}(x,\lambda)\,,\cr
x_0&= x+{x_1(x)\lambda\over\pi\epsilon}
+{x_2(x)\lambda^2\over8\pi^2\epsilon}+...\cr
&=xZ_x(x,\lambda)\,,
}\ee
where the dots stand for both the higher loop contributions and for the
higher order pole terms.
The unknown functions $\zeta_i(x)$ and $x_i(x)$ ($i=1$,$2$)
are determined from the following equations:
\be\label{renormeqs}
-\zeta_i(x)\tilde{\cal L}+{\pa\tilde{\cal L}\over\pa x}x_i(x)+
{\delta \tilde{\cal L}\over\delta\xi^k}\xi^k_i(\xi ,x)=\Sigma_i\,.
\ee
Eqs.~(\ref{renormeqs}) express the finiteness of the generalized quantum
effective action, $\Gamma(\xi)$, \cite{hulto} \cite{osb2}
to the corresponding order in perturbation
theory.
In Eqs.~(\ref{renormeqs}) $\xi^k_i(\xi ,x)$ may depend in an arbitrary way
on the parameter, $x$,
and on the fields, $\xi^j$, the only requirement being that $\xi^k_i(\xi ,x)$
contain no derivatives of $\xi^j$. This freedom
is related to the diffeomorphism
invariance of the renormalized theory.
Eqs.~(\ref{renormeqs}) admits a simple
interpretation: the general counterterms of the $\sigma$-models
may be absorbed by the  renormalization of the
coupling and the parameter(s) together with
a (in general non-linear) redefinition of the fields $\xi^j$:
\be\label{renormxi}
\xi^j_0=\xi^j+{\xi^j_1(\xi^k,x)\lambda\over\pi\epsilon}
+{\xi^j_2(\xi^k,x)\lambda^2\over8\pi^2\epsilon}+...
\ee
The functions $\xi^j_1$, $\xi^j_2$ are also determined
from Eqs.~(\ref{renormeqs}).
In the special case when $\xi_i^k$ depends
linearly on $\xi $ i.e.\  $\xi_i^k(\xi ,x)=\xi^k y_i^k(x)$,
Eqs.~(\ref{renormxi})
simplify to an ordinary multiplicative wave function renormalization.
We emphasize that it is not a priori guaranteed that Eqs.~(\ref{renormeqs})
may be solved at all
for the functions $\zeta_i(x)$, $x_i(x)$ and $\xi^k_i(\xi ,x)$.
If this happens to be the case then
the renormalization of the model is not possible within the
restricted subspace
characterized by the coupling(s) and the parameter(s)
in the (infinite dimensional)
space of  metrics and torsions.
This implies that
the model is not renormalizable in the ordinary, field theoretical
sense but only in the generalized sense \cite{friedan}, i.e.\ with an
infinite number of couplings.

\section{The deformed SU(2) principal $\sigma$-model and some of its duals}

Consider the following one parameter deformation of the
SU(2) principal $\sigma$-model Lagrangian:
\be\label{lagrsig}
{\cal L}=-{1\over2\lambda}\bigl(\sum\limits_{a=1}^3J_\mu^aJ^{\mu a}+
 gJ_\mu^3J^{\mu 3}\bigr)\,,
\ee
where $J_\mu=G^{-1}\pa_\mu G=J_\mu^a\tau^a$
where $\tau^a=\sigma^a/2$ and the $\sigma^a$ are the standard Pauli matrices,
with $G$ being an element
of SU(2) and $g$ is the parameter of the deformation.
From the Lagrangian (\ref{lagrsig})  it is clear that the
global $SU(2)_L\times SU(2)_R$
symmetry of the undeformed principal $\sigma$-model is broken to
$SU(2)_L\times U(1)_R$ by the $J_\mu^3J^{\mu 3}$ term.
Setting $g=0$
corresponds to the principal $\sigma$-model, while for $g=-1$ the $O(3)$
$\sigma$-model is obtained as can be seen from Eq.~(\ref{lagrsig2}) below.
In the following we
shall make perturbation in the coupling $\lambda$ while treating
$g$ as a parameter.
Using the Euler angles ($\phi$,$\theta$,$\psi$) to parametrize
the elements of $SU(2)$ $G$ is written as
\be\label{euler}
G=e^{i\phi\tau^3}e^{i\theta\tau^1} e^{i\psi\tau^3}\,.
\ee

Then the Lagrangian of the deformed model (\ref{lagrsig})  becomes
\be\label{lagrsig2}
\eqalign{
{\cal L}=&{1\over2\lambda}\bigl\{ (\pa_\mu\theta)^2+(\pa_\mu\phi)^2\bigl( 1+g
\cos^2\theta\bigr)+\cr
&(1+g)(\pa_\mu\psi)^2+2(1+g)\pa_\mu\phi\pa^\mu\psi\cos\theta\bigr\}\,.
}\ee
Clearly the deformed $\sigma$-model is a purely metric one.

Using the Killing vectors of the $SU(2)_L
\times U(1)_R$ symmetry and exploiting the manifest target space
covariance of the background field method one can prove that this model is
renormalizable in the ordinary sense: there is
no wave function renormalization for
$\theta$, $\phi$ and $\psi$, while the coupling constant
and the parameter get renormalized as in Eq.~(\ref{couplren}):
\be\label{couplren1}
\lambda_0=\mu^\epsilon Z_\lambda(\lambda ,g)\lambda ,\qquad
g_0=Z_g(\lambda ,g)g.
\ee
Both in the one and in the two loop orders
the residues of the single poles in $Z_\lambda(\lambda ,g)=
1+y_\lambda(\lambda ,g)/\epsilon +...$ and $Z_g(\lambda ,g)=
1+y_g(\lambda ,g)/\epsilon +...$ are determined from  Eqs.~(\ref{renormeqs}).
Though this yields five equations for the two unknown functions
$y_\lambda(\lambda ,g)$ and $y_g(\lambda ,g)$, since the model is
renormalizable these equations turn out to be  compatible and their
solution is given as:
\be\label{bf1}
\eqalign{
y_\lambda=&-{\lambda\over4\pi}\bigl( 1-g+{\lambda\over16\pi}(1-2g+5g^2)\bigr)\,
,\cr
y_g=&{\lambda\over2\pi}(1+g)\bigl( 1+{\lambda\over8\pi}(1-g)\bigr)\,.
}\ee
Let us next recall a useful relation between the
$\beta$ functions and the wavefunction renormalization in a theory
with two couplings.
Consider a theory with two couplings
(or one coupling and one parameter) denoted by $\alpha$ and $\gamma$, whose
bare and renormalized values are related as:
\be\label{couplren2}
\alpha_0=\mu^{a\epsilon}Z_\alpha(\alpha ,\gamma)\alpha\,,\qquad
\gamma_0=\mu^{b\epsilon}Z_\gamma(\alpha ,\gamma)\gamma\,.
\ee
Defining their $\beta$ functions in the standard way:
$\beta_\alpha=\mu{d\alpha\over d\mu}$,
$\beta_\gamma=\mu{d\gamma\over d\mu}$,
then these are determined by the
residues of the simple poles in the renormalization constants $Z_\alpha$
and $Z_\gamma$, $y_\alpha(\alpha ,\gamma)$ and
$y_\gamma(\alpha ,\gamma)$ as follows:
\be\label{beta1}
\eqalign{
\beta_\alpha=&\alpha\bigl( a\alpha{\pa y_\alpha\over\pa\alpha}+
b\gamma{\pa y_\alpha\over\pa\gamma}\bigr)\,, \cr
\beta_\gamma=&\gamma\bigl( b\gamma{\pa y_\gamma\over\pa\gamma}+
a\alpha{\pa y_\gamma\over\pa\alpha}\bigr)\,.
}\ee
From Eqs.~(\ref{bf1},\ref{beta1})
one obtains the $\beta$ functions of the deformed $\sigma$-model:
\be\label{betasig1}
\eqalign{
\beta
_\lambda=&-{\lambda^2 \over4\pi}\bigl( 1-g+{\lambda\over8\pi}(1-2g+5g^2)\bigr)\,
,\cr
\beta_g=&{\lambda\over2\pi}g(1+g)\bigl( 1+{\lambda\over4\pi}(1-g)\bigr)\,.
}\ee
It is easy to see, that the $g=0$ resp.\ the $g=-1$ lines are
fixed lines under
the renormalization group, and $\beta_\lambda$
reduces to the $\beta$
function of the principal $\sigma$-model, resp.\ of the $O(3)$
$\sigma$-model on them.
In the ($\lambda\ge0$, $g<0$) quarter of the ($\lambda$,$g$) plane
the renorm trajectories
run into $\lambda=0$, $g=-1$; while for $g>0$ they run to infinity. This
implies that the $g=0$ fixed line corresponding to the principal
$\sigma$-model is `unstable' under the deformation.

The Lagrangian of the deformed $\sigma$-model given by Eq.~(\ref{lagrsig2})
exhibits two
obvious Abelian isometries that can be used to construct two different
(Abelian) duals: namely the translations in the $\phi$ and $\psi$
fields; we call the models obtained this way the `$\phi$ dual' and the
`$\psi$ dual' of the deformed $\sigma$ model (\ref{lagrsig2}).
From Eq.~(\ref{euler}) it is clear that these translations correspond to
multiplying the $SU(2)$ element, $G$,
by a constant, diagonal $SU(2)$ matrix from the left (respectively
from the right). Since in the dual variables there is always a translational
symmetry and the duality transformation amounts to
a canonical transformation (classically) the `$\psi$ dual' is
expected to show the
full remaining $SU(2)\times U(1)$ symmetry of the original
model (\ref{lagrsig2}), while for the
$\phi$ dual only a $U(1)\times U(1)$ symmetry is expected.

\subsection{The `$\psi$ dual' model}
Let us start first with the `$\psi$ dual' of the deformed $\sigma$ model.
Its Lagrangian is easily found to be using
Buscher's formulae (\ref{dmetr}) and Eq.~(\ref{lagrsig2}):
\be\label{psilag}
\eqalign{
{\cal L}
=&{1\over2\tilde\lambda}\Bigl( (\pa_\mu\theta)^2+(\pa_\mu\phi)^2
\sin^2\theta+(\pa_\mu h)^2\cr
&+2a\cos\theta\epsilon^{\mu\nu}
\pa_\mu h\pa_\nu\phi\Bigr)\,,
}\ee
where $h$ denotes the (appropriately scaled) variable dual to $\psi$ and
($\tilde\lambda$,$\tilde g$) stand for the couplings of the dual model.
The couplings of the original (\ref{lagrsig2}) and of the dual
model (\ref{psilag}) are related (at the classical level) as
\be\label{couplrel}
\tilde\lambda =\lambda\quad a=\sqrt{1+\tilde g}
\quad \tilde g=g\,.
\ee
Note the
appearance of a non trivial torsion potential in Eq.~(\ref{psilag})
generated by the off diagonal $g_{\psi\phi}$ terms of the original
purely metric model Eq.~(\ref{lagrsig2}). For $a=0$ Eq.~(\ref{psilag})
reduces to the Lagrangian of the
$O(3)$ $\sigma$-model (apart from a decoupled free field), and it is easy to
show that for all values of $a$ it shows the expected
$SU(2)\times U(1)$ symmetry, indeed. For
$a=1$ the Lagrangian of the `$\psi$ dual'
becomes similar but not identical to that
of the so called `pseudo dual' of the $SU(2)$ principal model
\cite{zakmi}, \cite{nap}.
The difference between the `pseudo dual' model of Ref.~\cite{zakmi}
and the `$\psi$ dual' models is that the metric is
flat for the first model while it is not 
for the `$\psi$ dual' one.

The three Killing vectors generating
an $SU(2)$ of the (global) symmetry algebra of
(\ref{psilag}) act on a two-sphere, i.e.\ they are not linearly
independent. For this reason
the Killing equations expressing the
symmetry of the counterterms are not restrictive enough, so
from this $SU(2)$ symmetry alone one cannot
conclude
that the $\psi$ dual model is renormalizable in the ordinary sense.
Taking into account the additional restrictions following from the
discrete symmetry
$h,\theta ,\phi\rightarrow -h,-\theta ,-\phi$
of Eq.~(\ref{psilag}),
makes it possible to show that the complete renormalization of the model
amounts to just an ordinary multiplicative renormalization of
the couplings and of the $h$ field:
\be\label{psiren1}
\eqalign{
{\tilde\lambda}_0=&\mu^\epsilon Z_{\tilde\lambda}({\tilde\lambda} ,a)
{\tilde\lambda} ,\quad
a_0=Z_a({\tilde\lambda} ,a)a\,,\cr
 h_0=&hZ_h({\tilde\lambda} ,a)\,.
}\ee
The one and two loop counterterms
$\Sigma_1$, $\Sigma_2$
 are relatively simple expressions:
\be\label{psictms1}
\eqalign{
\Sigma_1=&{1\over4}\Bigl( (1-{a^2\over2})\Omega
-{a^2\over2}(\pa_\mu h)^2\Bigr),\cr
\Sigma_2=&{1\over8}\Bigl( ({3a^4\over2}+4(1-a^2))\Omega
+{3a^4\over2}(\pa_\mu h)^2\Bigr)\,,
}\ee
where $\Omega=(\pa_\mu\theta)^2+(\pa_\mu\phi)^2\sin^2\theta$.
A straightforward computation based on Eq.~(\ref{renormeqs}) yields
for the simple pole
parts of $Z_j({\tilde\lambda} ,a)=
1+y_j({\tilde\lambda} ,a)/\epsilon +...$, $j={\tilde\lambda} ,a,h$:
\be\label{psictms2}
\eqalign{
y_{\tilde\lambda}=&-{\tilde\lambda\over2\pi}(1-{a^2\over2})-
{\tilde\lambda^2\over8\pi^2}
\bigl({3a^4\over8}+1-a^2 \bigr),\cr
y_a=&-{\tilde\lambda\over4\pi}(1-a^2)-{\tilde\lambda^2\over16\pi^2}
\bigl({3a^4\over4}+1-a^2 \bigr),\cr
y_h=&-{\tilde\lambda\over4\pi}-{\tilde\lambda^2\over16\pi^2}(1-a^2)\,.
}\ee
Using now Eq.~(\ref{beta1}) leads immediately to the $\beta$ functions:
\be\label{psibetaf}
\eqalign{
\beta_{\tilde\lambda}=&-{{\tilde\lambda}^2\over2\pi}(1-{a^2\over2})-
{{\tilde\lambda}^3\over4\pi^2}
\bigl({3a^4\over8}+1-a^2 \bigr)\,,\cr
\beta_a=&-{\tilde\lambda a\over4\pi}(1-a^2)-{{\tilde\lambda}^2a\over8\pi^2}
\bigl({3a^4\over4}+1-a^2 \bigr) \,.
}\ee
For convenience let us give here also $\beta_{\tilde\lambda}$ expressed
in terms of ${\tilde g}$ together with
$\beta_{\tilde g}=2a\beta_a$:
\be\label{psibetaf1}
\eqalign{
\beta_{\tilde\lambda}=&-{{\tilde\lambda}^2 \over4\pi}\bigl( 1-{\tilde g}+{{\tilde\lambda}\over8\pi}
(3-2{\tilde g}+3{\tilde g}^2)\bigr)\,,\cr
\beta_{\tilde g}=
&{{\tilde\lambda}\over2\pi}(1+{\tilde g})\bigl( {\tilde g}-{{\tilde\lambda}\over8\pi}(3+2{\tilde g}+3{\tilde g}^2)\bigr)\,.
}\ee
These $\beta$ functions show a number of interesting properties.
First of
all in the one loop order they are completely equivalent to those of
Eq.~(\ref{betasig1}) taking into account the relation (\ref{couplrel})
between
the couplings ($\lambda$,$g$) and ($\tilde\lambda$,$a$) i.e.\ 
$g=a^2-1$, which holds true at the classical level.
This equivalence is
seemingly broken, however, at the two loop level, that is
to that order the two sets of $\beta$ functions,
Eqs.~(\ref{betasig1}) and (\ref{psibetaf1}) are different.
Nevertheless this does not imply the inequivalence of the
two models in perturbation theory since
in the case of more than one coupling
the two loop $\beta$ functions are already scheme dependent quantities.
Indeed the perturbative redefinition of the couplings
(a change of scheme)
\be
\tilde\lambda=\lambda +{\lambda^2\over4\pi}F(g)\quad
\tilde g=g+{\lambda\over4\pi} H(g)\,,
\ee
can be easily seen to change the two loop coefficients of the
$\beta$ functions (\ref{psibetaf1}).
In fact one can simply try to determine the arbitrary
functions $F(g)$, $H(g)$, so that the $\beta$ functions of
the original and the dual model agree up to this order.
A direct calculation to equate the corresponding $\beta$ functions
yields an explicitly solvable system of
two first order differential equations for $F(g)$ and $H(g)$.
The general solution contains of course two constants of
integration which can only be determined from some other arguments
(regularity in $g$ and the constraint coming from self-dual point
at $g=-1$).
The following change of scheme (transforming Eqs.~(\ref{betasig1}),
(\ref{psibetaf1}) into each other up to second order in $\lambda$
by construction) determined from the two previous arguments are
\be\label{psicoupl}
\tilde\lambda=\lambda+{\lambda^2\over4\pi}(1+g)\,,\quad
\tilde g=g+{\lambda\over4\pi}(1+g)^2\,.
\ee
To really establish the perturbative equivalence of (\ref{lagrsig2})
(\ref{psilag}) up to two loops
one has to make a more direct comparison between some
`physical' quantities in the two models.
As one cannot directly relate operators under the T-duality
transformation, we have chosen to compare the
$<\theta(k_1) \theta(k_2) \phi(k_3) \phi(k_4)>$ four point functions
computed both in the original
(\ref{lagrsig2}) and in the dual model (\ref{psilag}). Clearly the
two four point functions must agree as $\theta$ and $\phi$ were
not even touched upon. As already alluded to one has every right to
expect that the T-duality transformation together with the use of
dimensional regularization corresponds to
a change of scheme.
Then by comparing some `physical' quantities in the two
different looking theories corresponding to a change of scheme,
one directly obtains the relation between them.
It is of course clear that such four point
functions are not really physical as they are e.g.\ coordinate dependent.
For relating the couplings ($\lambda$,$g$) and
($\tilde\lambda$,$\tilde g$) they are, however, perfectly well suited.
Furthermore by computing
$<\theta(k_1) \theta(k_2) \phi(k_3) \phi(k_4)>$ up to only one loop order,
one can check the two loop relation between the couplings!
The reason for this `miracle' is simply that the tree level
amplitudes in (\ref{lagrsig2}) and (\ref{psilag})
are already proportional to $\lambda$.
It is worth pointing out that the calculation
of $<\theta(k_1) \theta(k_2) \phi(k_3) \phi(k_4)>$ provides
a nontrivial cross check on the two loop $\sigma$-model counterterms
(\ref{lag12}).
The computation of these four point functions is a
straightforward though a somewhat tedious exercise in perturbation
theory and we omit the details here.
The final outcome of the calculation is precisely the previously deduced 
relation, Eq.~(\ref{psicoupl}), between the two sets of couplings.
This is in agreement with our expectation that
when $g_{00}$ is constant the `naive' T-duality transformation
 yields an equivalent model.

An alternative way to express the correspondence between the two
models is to get rid of
the scheme dependence of the $\beta$ functions
by eliminating one of the parameters ($g$ or $a$)
in $\beta_\lambda$ in favour of a
{\sl renormalization group invariant parameter} ($M$ resp.\
$\tilde M$).
A straightforward computation yields,
that the invariant parameter characterizing the trajectories under
the renormalization group equations (\ref{betasig1}) has the form:
\be\label{inv1}
M=-{\lambda^2g\over (1+g)^2}-{\lambda^3g\over4\pi (1+g)}\,,
\ee
while for Eqs.~(\ref{psibetaf1}) it is given by
\be\label{inv2}
\tilde M=-{\tilde\lambda}^2{a^2-1\over a^4}+{{\tilde\lambda}^3\over4\pi a^2}\,.
\ee
(The signs have been chosen here to guarantee that $M$
($\tilde M$) $>0$ in the
most interesting domain $\lambda >0$, $0>g>-1$, ($1>a>0$)). If
$M\neq 0$ ($\tilde M\neq 0$), then, expressing perturbatively $g$
(respectively $a$) from Eq.~(\ref{inv1}) (resp.\ Eq.~(\ref{inv2}) yields
\be\label{inv3}
g(\lambda,M)=-1+{\lambda\over\sqrt{M}}\,,\quad
a^2(\tilde\lambda,\tilde M)={\tilde\lambda\over\sqrt{\tilde M}}\,,
\ee
and using them to compute $\beta_\lambda$ in Eqs.~(\ref{betasig1})
(resp.\ $\beta_{\tilde\lambda}$ in Eqs.~(\ref{psibetaf1}) ) shows that the
two expressions become identical
provided $M=\tilde M$.
If $M=0$, then Eq.~(\ref{inv1}) immediately
yields $g=0$ (implying that this case is
the $SU(2)$ principal model), while for $\tilde M=0$
 from Eq.~(\ref{inv2}) one
obtains $a^2=1+{\tilde\lambda\over4\pi}$, and after eliminating it
from Eqs.~(\ref{psibetaf}) $\beta_{\tilde\lambda}$ becomes identical
with that of the $\beta$ function of the
principal model.
Of course all these findings are compatible with the deformed
$\sigma$-model being two loop quantum equivalent to its `$\psi$ dual'.

\subsection{The `$\phi$ dual' model}
The Lagrangian of the $\phi$ dual of the deformed $\sigma$-model
has the form:
\be
\label{philag}
\eqalign{
{\cal L}
=&{1\over2\lambda \Theta}\bigl[ (\pa_\mu f)^2
+(1+g)\sin^2\theta(\pa_\mu\psi)^2\cr
&+\Theta(\pa_\mu\theta)^2 +2(1+g)\cos\theta\epsilon^{\mu\nu}
\pa_\mu f\pa_\nu\psi\bigr],
}\ee
where $\Theta=1+g\cos^2\theta$ and
$f$ denotes the  variable dual to $\phi$. Setting $g=0$ in
Eq.~(\ref{philag})
gives the same `pseudo dual'-- like Lagrangian as in the case of the $\psi$
dual, while for $g=-1$ one obtains a model resembling the $O(3)$
$\sigma$-model.
From the $SU(2)\times U(1)$ symmetry of the original model
(\ref{lagrsig2}) only a $U(1)\times U(1)$ remains manifest
in Eq.~(\ref{philag}) 
(the two translational symmetries in the variables $f$ and $\psi$).
Since this $U(1)\times U(1)$ symmetry is not restrictive enough
(the only constraint on the counterterms coming from it
is that they can only depend on the $\theta$ field), to 
prove renormalizability of (\ref{philag}) in
the restricted sense. Computing, nevertheless, $\Sigma_1$ and
$\Sigma_2$ reveals that up to this order the structure of the Lagrangian
Eq.~(\ref{philag}) is preserved: both in the metric and in
the torsion potential
only the non-vanishing elements receive corrections, while the vanishing
elements do not. The explicit form of $\Sigma_1$ is:
\be
\label{phictms}
\eqalign{
\Sigma_1 =&-{1\over8\Theta^3}\bigl[
-8g_{\scr{+}}z\sin^2\theta\epsilon^{\mu\nu}\pa_\mu f\pa_\nu\psi\cr
 &-(1+3g-6gg_{\scr{+}}z^2+g^2g_{\scr{-}}z^4)\Theta(\pa_\mu \theta)^2\cr
&-g_{\scr{+}}^2\sin^2\theta(1-4gz^2-g^2z^4)(\pa_\mu \psi)^2\cr
&+(g_{\scr{-}}(1-g^2z^4)+4gg_{\scr{+}}z^2)(\pa_\mu f)^2 \bigr]\,,
}\ee
where $z\equiv\cos\theta$, $g_{\scr{+}}\equiv1+g$ and
$g_{\scr{-}}\equiv1-g$.
Let us now try to convert this one loop counterterm into a coupling
and parameter renormalization as in Eqs.~(\ref{couplren}), accompanied by a
non-linear redefiniton
of $\theta$ together with some multiplicative renormalization of the
$\psi$ and $f$ fields:
\be\label{phicpren}
\eqalign{
&\lambda_0=\mu^\epsilon\lambda\Bigl( 1+{\zeta_1(g)\lambda\over\pi\epsilon}
\Bigr),\
g_0= g+{g_1(g)\lambda\over\pi\epsilon}\,,\cr
&f_0=f\bigl( 1+{y^{(1)}_f(g)\lambda\over\pi\epsilon}\bigr)\,,\cr
&\theta_0=\theta+{T_1(\theta ,g)\lambda\over\pi\epsilon}\,,\
\psi_0= \psi\bigl( 1+{y^{(1)}_p(g)\lambda\over\pi\epsilon}\bigr)\,.
}\ee
Eq.~(\ref{renormeqs}) yields four equations
(corresponding to the four non vanishing
elements of the metric and the torsion potential)
for the five unknown functions in Eqs.~(\ref{phicpren}) with
only one 
depending on two variables ($g$, $\theta$). 
Therefore it is by no
means obvious that this problem has a solution at all. This is especially so,
since equating the coefficients of $(\pa_\mu\theta)^2$ on the two sides of
Eq.~(\ref{renormeqs}) yields a differential equation for
$T_1(\theta,g)$ from which we have found
\be\label{phiren2}
T_1(\theta ,g)=-{g\cos\theta\sin\theta\over2\Theta}\,,\quad 
\zeta_1(g)=-{g_{\scr{-}}\over4}\,.
\ee
Thus we have three functions of one variable ($g$) at our disposal
to satisfy the
three remaining equations in two variables ($\theta$ and $g$).
Nevertheless, after some effort one finds that choosing
\be\label{phiren3}
g_1(g)={gg_{\scr{+}}\over2}\,,\ y^{(1)}_f(g)=-{g_{\scr{-}}\over4}\,,\ 
 y^{(1)}_p(g)=0\,,
\ee
guarantees that all equations are satisfied.
Therefore in the one loop order
the `$\phi$ dual' model is also renormalizable in the restricted sense.
Furthermore
extracting the residues of the simple poles of $Z_\lambda$ and $Z_g$ from
Eqs.~(\ref{phicpren}, \ref{phiren3}) shows that up to this order
the $\beta$ functions of (\ref{philag})
are just identical
to that of the deformed $\sigma$-model 
in complete agreement with their one loop equivalence.

At the two loop order it still remains true that in (\ref{philag})
only the
non vanishing metric and torsion potential terms receive corrections.
As the explicit form of the two loop counterm, $\Sigma_2$, is rather
complicated we do not display it here.
$\Sigma_2$ is again a
rational function of $\cos\theta$ similarly to $\Sigma_1$.
There is a dramatic change, however, as compared to the previous
(one loop) case when we try to determine from $\Sigma_2$ 
the renormalization of the couplings together with
the same type of field renormalizations as in Eqs.~(\ref{phicpren}).
Integrating
 the differential equation for $T_2(\theta,g)$ yields namely
\be\label{phi2loop}
\eqalign{ 2&T_2(\theta ,g)=\left[\zeta_2+{3g_{\scr{-}}^2+
4g\over8}\right]\theta
 +{(gz)^3\sin\theta\over \Theta^2}\cr
+&C-{g\over2g_{\scr{+}}}\Bigl[{z\sin\theta\over\Theta}
-{1\over\sqrt{g_{\scr{+}}}}{\rm arctg}(\sqrt{g_{\scr{+}}}\ {\rm cotg}\theta)
\Bigr]\,,
}\ee
where $C$ is the constant of integration.
The problem is now that no choice of $\zeta_2(g)$
and $C$ in Eq.~(\ref{phi2loop}) could make $T_2(\theta,g)$
a purely rational expression of $\cos\theta$ and $\sin\theta$.
Recalling that $\Sigma_2$ is a rational expression
in $\cos\theta$ and looking at the remaining three
equations for $g_2(g)$,
$y^{(2)}_f(g)$ and $y^{(2)}_p(g)$ it is not difficult to see
that there is no way to satisfy Eqs.~(\ref{renormeqs})
in the two loop order.
The implication of this result is that in the 
two loop order in perturbation theory,
the `$\phi$ dual' model is not
renormalizable in the restricted (field theoretical) sense.
This clearly shows that application of the {\sl classical} T-duality
transformations (\ref{dmetr}) in the standard perturbative
$\sigma$-model renormalization framework 
may lead to inequivalent dual models. As we have already mentioned
the quantum inequivalence of the dual model
{\sl defined by the classical} Buscher formulae (\ref{dmetr})
is expected to be related
to the a presence of a nonconstant $g_{00}$, hence to renormalization
effects.

\section{The non Abelian dual of the principal $\sigma$-model}

In this section we investigate the problem of perturbative quantum
equivalence of the principal $\sigma$-model with its non Abelian
dual \cite{freeto}, \cite{fj}, \cite{frats}, \cite{curza}.
The non Abelian dual of the principal $\sigma$-model can be deduced
in a way similar to that of the Abelian T-duality transformation,
by making some formal manipulations in the partition function. 
One can start for example with the 2d Freedman-Townsend model
\be\label{ftlagr}
{\cal L}_{\scr{\rm FT}}=B^a\epsilon^{\mu\nu}F_{\mu\nu}^a
+A_\mu^aA^{a\mu}\,,
\ee
where $a=1,2,\dots$ denotes the Lie algebra (semi simple) indices, 
$F_{\mu\nu}^a$ is the standard field-strength
tensor of the non Abelian vector fields, $A_\mu^a$,
and $B^a$ are auxiliary fields (Lagrange
multipliers enforcing the vanishing of the field tensor).
By integrating over the auxiliary fields one obtains the principal
$\sigma$-model while integration over $A^a_\mu$
yields the corresponding non Abelian dual.
An alternative derivation of the non Abelian duality generalizing
the gauging procedure of \cite{rove} for the Abelian duality
was given in \cite{osqu}.

In what follows we shall only consider the simplest case,
namely the Lie algebra being $SU(2)$.
Let us present first a `universal' Lagrangian containg both the
principal $\sigma$-model and its non Abelian dual: 
\be\label{nalagr}
\eqalign{
{\cal L}={1\over 2e^2}\big\{& \partial_\mu r\partial^\mu r+A(r)
\partial_\mu n^a\partial^\mu n^a\cr
&+B(r)\epsilon^{abc}\epsilon_{\mu\nu}
n^a\partial_\mu n^b \partial_\nu n^c\big\}\,,
}\ee
where an element of $SU(2)$ has been parametrized by  
a unit vector, $n^a$, and by a radial variable $r$.
The principal $\sigma$-model corresponds to the choice
$A(r)=\sin^2r$ and $B(r)\equiv 0$ 
while for the non Abelian dual $A(r)=r^2/r_{\scr{+}}$ and
$B(r)=2r^3/r_{\scr{+}}$, where $r_{\scr{+}}=1+4r^2$.
To make contact between the present formulation
of the principal model
and its previously given Lagrangian (\ref{lagrsig}) (for $g=0$)
we note that (\ref{nalagr})
corresponds to parametrizing $G$ in (\ref{lagrsig}) as
$G=\cos r+i\sin r n^a\sigma^a$ (and writing $\lambda=2e^2$).

After the results of the previous section it is natural to guess
that troubles are likely to arise when checking
the quantum equivalence between the principal $\sigma$-model and
its `naive' (i.e.~classical) non Abelian dual.
The one loop agreement of the corresponding $\beta$ functions
has been established long ago in Ref.~\cite{fj} and
in Ref.~\cite{frats} it has been even argued that the $\beta$ functions
are completely equivalent. It will be demonstrated below, however,
that in analogy to the `$\phi$ dual' studied in the previous section,
the `naive' non Abelian dual is simply not renormalizable with a
single coupling at the two loop level, while the $\beta$ functions 
in the two models
agree up to one loop order, indeed. 

From Eq.~(\ref{nalagr}) it is clear that both the original and the
dual model have a manifest $SU(2)$ symmetry.
Eq.~(\ref{lagrsig}) for $g=0$ clearly shows the full $SU(2)\times SU(2)$
symmetry of the principal $\sigma$-model.
It is well known of course that the principal $\sigma$-model
is renormalizable in the restricted sense. This fact
is due to its high degree of symmetry, in particular the
corresponding (3d) metric in (\ref{nalagr}) possesses 
three linearly independent Killing vectors. This restricts the counterterms
so strongly that the standard multiplicative renormalization is sufficient
to remove the divergences.
For the non Abelian dual model where the Killing vectors of
the single $SU(2)$ symmetry act on a
two sphere (that only two of them is linearly independent)
this symmetry is not restrictive enough to establish 
its renormalizability.  

The one and two loop counterterms for both cases can be written in a
unified way:
\be\label{nactrms}
\eqalign{
\Sigma_i=N_i\big\{& Z_i(r)\partial_\mu r\partial^\mu r+A_i(r)
\partial_\mu n^a\partial^\mu n^a\cr
&+B_i(r)\epsilon^{abc}\epsilon_{\mu\nu}
n^a\partial_\mu n^b \partial_\nu n^c\Big\}\,,
}\ee
where $i=1,2$ and $N_1=1/4$ and $N_2=1/8$. This form of the counterterms
agrees with the known property of dimensional regularization of
preserving the global symmetries of the target manifold.
To respect the manifest $SU(2)$ symmetry of (\ref{nactrms})
when discussing the nonlinear field renormalizations, the only thing
one can
allow for is a redefinition of the radial variable, $r$. Therefore
when the models are renormalizable the counterterms should just give
the renormalization of the (single) coupling and the $r$ field:
\be\label{nacoupls}
\eqalign{
e_0^2=&\mu^{\epsilon }e^2\Bigl(1+{\zeta_1e^2\over\pi\epsilon}+
{\zeta_2e^4\over8\pi^2\epsilon}\Bigr)=\mu^{\epsilon }e^2Z_{e^2}\,,\cr 
r_0=&r+{r_1(r)e^2\over\pi\epsilon}+
{r_2(r)e^4\over8\pi^2\epsilon}\,.}
\ee
In the case of the principal $\sigma$-model the actual expressions for
$A_i$, $B_i$ and $Z_i$ are given as:
\be\label{nactrms2}
\eqalign{
A_i(r)=&\delta_iA(r)\,,\cr
\delta_1=&Z_1=1\,,}
\qquad
\eqalign{
 B_i(r)\equiv& 0\,,\cr
\delta_2=&Z_2=2\,.}
\ee
Using now (\ref{nactrms2}) 
in Eqs.~(\ref{renormeqs}) yields that $r_1(r)\equiv 0$,
 $r_2(r)\equiv 0$ and $Z_{e^2}=1-
{e^2/\pi\epsilon}-{e^4/4\pi^2\epsilon}$.
Then it is easy reproduce the known result for
the two loop $\beta$ function of the principal $\sigma$-model:
\be
\beta_{e^2}=-{e^4\over\pi}\left(1+{e^2\over2\pi}\right)\,.
\ee
In the case of the non Abelian dual the $A_i$, $B_i$ and $Z_i$ functions are
somewhat more complicated:
\be\label{nactrms3}
\eqalign{
A_1(r)=&{r^2(r_{\scr{+}}^2-2r_{\scr{-}})\over r_{\scr{+}}^3}\,,\ 
Z_1(r)=-{r_{\scr{+}}^2+4r_{\scr{-}}\over r_{\scr{+}}^3}\,,\cr
B_1(r)=&{8r^3\over r_{\scr{+}}^3}\,,\ 
A_2(r)={2r^2(3r_{\scr{+}}^4-80r_{\scr{+}}+32)\over r_{\scr{+}}^5}\,,\cr
B_2(r)=&-{64r^3(r_{\scr{+}}^2+4r_{\scr{+}}-2)\over r_{\scr{+}}^5}\,,\cr 
Z_2(r)=&{6(r_{\scr{+}}^3+8(r_{\scr{+}}^2+r_{\scr{-}}-2))\over r_{\scr{+}}^3}\,,
}\ee
where we have introduced $r_{\scr{-}}=4r^2-1$.
Eqs.~(\ref{renormeqs}) lead to three equations for the two unknown
functions $r_i$, $\zeta_i$ (coming from
equating the coefficients of the three different tensor structures
arising on the two sides):
\be\label{naeqs}
\eqalign{
-\zeta_if(r)+&r_i(r){\pa f\over\pa r}(r)
-f_i(r)=0,\cr
-\zeta_i+&2{d r_i\over d r}(r)-Z_i(r)=0\,,}
\ee
where $f=A$, $B$.
The first two equations in (\ref{naeqs}) can be solved algebraically for 
$\zeta_i$ and $r_i(r)$; defining $b_i={A_i(r)/A(r)}$
and $c_i={B_i(r)/rA(r)}$ the solutions are simply 
\be\label{nares}
\zeta_i={c_i-(3+4r^2)b_i\over r_{\scr{+}}},\quad
r_i(r)=r({c_i\over2}-b_i). 
\ee
Clearly, to be self-consistent, $\zeta_i$ must be independent of $r$  
and $r_i(r)$ must still
satisfy the third equation in (\ref{naeqs}).
From the given expressions for $A_i$, $B_i$ and $Z_i$ in Eqs~(\ref{nactrms3})
one finds that at one loop $\zeta_1=-1$
and $r_1(r)=-rr_{\scr{-}}/r_{\scr{+}}$,
which indeed solves the third equation in (\ref{naeqs}).
Then the one loop 
$\beta$ function of the principal $\sigma$-model can be immediately
reproduced indicating the equivalence of the two theories to
this order.
In the
two loop order, however, we find that $\zeta_2$ is not $r$ independent, 
thus the non Abelian dual is not renormalizable!
This shows that the `naive' version of the non Abelian duality
transformation as it stands in Eq.~(\ref{nalagr}) yields 
an inequivalent model at the two loop level in perturbation theory.
This last conclusion has also been reached 
in Ref.~\cite{subtu}, we disagree, however, with some of the other
claims of that paper.

Let us remark that in all of the examples 
investigated so far we have also checked that our conclusions
on the two loop 
(non)renormalizability of the models in question is independent of the well
known ambiguity (or freedom) in the counterterms
(corresponding to target space diffeomorphism invariance)  
\cite{hulto}, \cite{metse}, \cite{osb1}, \cite{howpa}, \cite{osb2}.
In all of the cases we have found
that the appropriate covariant derivatives of $H^2=H_{ijk}H^{ijk}$ either
vanish identically or give no contribution to the consistency 
conditions like the one in Eq.~(\ref{nares}).

\section{Conclusions}
We have shown on a number of examples that the `naive' (tree level) T-duality
transformations in 2d $\sigma$-models cannot be
exact symmetries of the quantum theory. The `naive' Abelian duality
transformations Eqs.~(\ref{dmetr})
are correct to one loop in perturbation theory, they break down in general,
however, at the two loop order.
This two loop problem is expected to be connected to regularization
issues in the
functional integral when deriving the duality transformation formulae
Eqs.~(\ref{dmetr}) \cite{tse1}.
In the very simple case of the
Abelian duality transformations (\ref{dmetr}) with $g_{00}$ {\sl constant}
when the derivation amounts to just a standard gaussian integration,
no problems are expected with the quantum equivalence of the dual theory
(`$\psi$-dual' model). We have found that in this case the dual model
is indeed equivalent to two loops to the original one, however,
there is a nontrivial
change of scheme involved when insisting on dimensional regularization.
That the `$\psi$-dual' model corresponds to changing the regularization
scheme in the original theory has also been checked by computing a suitable
four point function.

Our conclusion from the above (somewhat discouraging) results concerning
T-duality at the quantum level is certainly {\sl not} that 
there is no T-duality symmetry.
In the simplest conceivable case of two free fields, when simililar
problems arise at the two loop level, we have found a suitable 
modification of the duality formulae to ensure quantum equivalence
to two loops \cite{bfhp}. Encouraged by this positive result
we do expect that a suitable modification of the
`naive' duality transformations
is possible in general order by order in perturbation theory making
(Abelian) T-duality a true {\sl quantum} symmetry.

\section*{Acknowledgements}
One of us (P.~F.) would like to thank the organizers of
the 29$^{\rm th}$ Buckow Symposium, in particular
D.~L\"ust and G.~Weigt,
for the invitation to a very stimulating Conference
and for giving the opportunity to present the above results.

\end{document}